\documentclass[twocolumn]{bmcart}

\usepackage[utf8]{inputenc} 
\usepackage{graphicx}
\usepackage{rotating}
\usepackage{hyperref}
\usepackage{xcolor,colortbl}

\definecolor{Gray}{gray}{0.9}
\definecolor{LightCyan}{rgb}{0.88,1,1}
\newcolumntype{a}{>{\columncolor{Gray}}c}
\newcolumntype{b}{>{\columncolor{white}}c}

\startlocaldefs
\endlocaldefs

\begin{document}

\begin{frontmatter}

\begin{fmbox}
\dochead{Computational Virology - Parameter Estimation}


\title{Estimating viral infection parameters using Markov Chain Monte Carlo simulations}


\author[
   addressref={aff1},                   
   corref={aff1},                       
   noteref={n1},                        
   email={valeriu.predoi@astro.cf.ac.uk}   
]{\inits{VP}\fnm{Valeriu} \snm{Predoi}}


\address[id=aff1]{
  \orgname{Physics and Astronomy, Cardiff University}, 
  \street{The Parade},                     %
  \postcode{CF24 3AA}                                
  \city{Cardiff},                              
  \cny{United Kingdom}                                    
}


\begin{artnotes}
\note[id=n1]{Equal contributor} 
\end{artnotes}

\end{fmbox}


\begin{abstractbox}

\begin{abstract} 
Given a mathematical model quantifying the viral infection of pandemic influenza H1N1pdm09-H275 wild type (WT) and H1N1pdm09-H275Y mutant (MUT) strains, we describe a simple method of estimating the model's constant parameters using Monte Carlo methods. Monte Carlo parameter estimation methods present certain advantages over the bootstrapping methods previously used in such studies: the result comprises actual parameter distributions (posteriors) that can be used to compare different viral strains; the recovered parameter  distributions offer an exact method to compute credible intervals (similar to the frequentist 95\% parametric confidence intervals (CI)), that, in turn, using a suitable analysis statistic, will be narrower than the ones obtained from bootstrapping; given an appropriate computational parallelization, Monte Carlo methods are also faster and less computationally intensive than bootstrapping. We fit Gaussian distributions to the parameter posterior distributions and use a two-sided Kolmogorov-Smirnoff test to compare the two strains from a parametric point of view; our example result shows that the two strains are 94\% different. Furthermore, based on the obtained parameter values, we estimate the reproductive number $R_0$ for each strain and show that the infectivity of the mutant strain is larger than the wild type strain.

\end{abstract}




\end{abstractbox}
%

\end{frontmatter}



\section{Introduction}
\vspace{0.3cm}

The World Health Organization (WHO) declared the first influenza pandemic of the 21$^{\mathrm{st}}$ century in 2009 and described the virus (H1N1pdm09) as naturally resistant to adamantanes but susceptible to the neuraminidase (NA) inhibitors oseltamivir and zanamivir \cite{abedy,cent}. The H275$Y$ mutation within the NA gene was reported to be associated with the drug resistance over the past three years although the overall level of resistance has remained relatively low  The pandemic strain completely displaced the prior seasonal H1N1 strain (A/Brisbane/59/2007) \cite{abedy}, which, in the 2008-2009 season, was nearly 100\% resistant to oseltamivir \cite{dran}. It was initially thought that the mutation usually compromised strain fitness (30), therefore the dominance of an oseltamivir-resistant strain is rather surprising. A return to widespread oseltamivir resistance, with a mutated H1N1pdm09 virus, could have significant public health consequences \cite{abedy}.

The H275Y amino acid substitution of the neuraminidase gene is the most common mutation conferring oseltamivir resistance in the N1 subtype of the influenza virus. Using a mathematical model to analyze a set of in vitro experiments that allow for the full characterization of the viral replication cycle, \cite{holder} show that the primary effects of the H275Y substitution on the pandemic H1N1 (H1N1pdm09) strain are to lengthen the mean eclipse phase of infected cells (from 6.6 to 9.1 h) and decrease (by 7--fold) the viral burst size, $i.e.$ the total number of virions produced per cell; \cite{holder} also find, however, that the infectious--unit--to--particle ratio of the H275Y mutant strain is 12-fold higher than that of the oseltamivir-susceptible strain (0.19 versus 0.016 per RNA copy). The multicompartment mathematical model presented in \cite{holder} makes use of ordinary differential equations (ODE) to describe the virus as a dynamical system that undergoes different phases of evolution; the model is characterised by a set of model parameters $\theta_i$ that \cite{holder} estimate using bootstrapping. We will use the same mathematical model approach (albeit with a changed set of ODEs) and data set used in \cite{holder} but employ a different parameter estimation method -- we will use a set of Markov Chain Monte Carlo (MCMC) codes to robustly estimate $\theta_i$ parameter values and credible intervals from posterior distributions obtained from the MCMC simulations. Our results, combined with a minimal number of prior assumptions about the data, provide a more precise set of values for the viral parameters $\theta_i$ and the method could be standardized and used in the future as a stand-alone parameter estimation package for viral infectious disease modelling purposes. 

The primary aim of this work is to show the benefits of using MCMC parameter estimation techniques over the traditional bootstrapping methods presented in, e.g. \cite{holder}; secondary aims, connected to the primary, are to introduce the reader to the theoretical support of the analysis method, present a number of steps taken to optimize the MCMC analysis process (resolving computational issues with the numerical ODE solver, changing the ODE model to a more computationally-robust form) and to re-state and refine the results presented in \cite{holder} as obtained from using a different and more precise analysis method. As a whole, this work presents an easily standardizable analysis method for any given biological data set that can be described by a mathematical model whose parameters need to be estimated robustly and fast; the theoretical framewrok is mostly Bayesian - with the only exception of the use of a frequentist test comparing parametric distributions.

This article is divided as follows: in Section \ref{theory} we will present the theoretical fundamentals that we use throughout -- the mathematical model that describes the viral infection dynamics together with its parameters are described in \cite{holder}, what we will outline is the Bayesian support of the MCMC method and the statistic used for the proposed method of estimation; in Section \ref{computation} we describe the computational challenges poised by the MCMC method -- corrections applied to the numerical integration of the model's ODE set and a rewritten, more computationally-robust, ODE model; in Section \ref{results} we present the results of the study and formulate a discussion around the main concluding points, presented in the last section, Section \ref{conclusions}. 
\section{Theoretical aspects}
\label{theory}
\vspace{0.3cm}

A virus undergoes a number of phases in its trajectory towards infection and replication: (in brief) it will attach and enter the host cell, inject its genetic material and convert the cell to become a virus producer and then it will exit the host and repeat the cycle \cite{nowak}; these phases can be modelled mathematically looking at the virus and host cells as two distinct populations. Mathematical models have been applied to simulate viral disease spread characteristics before \cite{baccam,holder,nowak} -- approximating the virus with a dynamical system that can be characterised by a set of state variables and a set of parameters, an ODE system will quantify the relation between the state variables and their rate of change with time. For this work in particular, viral yield measurements may be simulated by using the following multicompartment ODE model, as described in \cite{holder}:
\begin{eqnarray}
\dot V_\mathrm{PFU} &=& \rho p\left(\sum_{j=1}^{n_I}I_j\right) - (c+c_\mathrm{RNA})V_\mathrm{PFU} \nonumber \\
\dot V_\mathrm{RNA} &=& p\left(\sum_{j=1}^{n_I}I_j\right) - c_\mathrm{RNA}V_\mathrm{RNA} \nonumber \\
\dot T &=& -\beta T V_\mathrm{PFU} \nonumber \\
\dot E_1 &=& \beta TV_\mathrm{PFU}-kE_0,~\mathrm{where}~k = n_E/\tau_E \nonumber \\
\dot E_i &=& -k\Delta E \nonumber \\
\dot I_1 &=& kE_{n_E}-\delta I_0,~\mathrm{where}~\delta = n_I/\tau_I \nonumber \\
\dot I_j &=& -\delta \Delta I \nonumber \\
\dot D &=& -\delta I_{n_I}D
\label{re0}
\end{eqnarray}
with $E_0$ and $I_0$ the values of $E$ and $I$ in the first eclipse and infectious compartments respectively and $\Delta E=E_{i} - E_{i-1}$ and $\Delta I=I_{j} - I_{j-1}$ and that describes the infection of a population of $N$ susceptible target cells $T$ at a rate $\beta V_\mathrm{PFU}$, here $V_\mathrm{PFU}$ is the quantity of infectious virus. Newly infected cells first undergo an eclipse phase $E$ of average duration $\tau_E$ before becoming infectious $I$ and producing virus at a constant rate $p$ for an average time $\tau_I$; $V_\mathrm{RNA}$ represents the total (number of RNA copies/ml) virus concentration, controlled by the virus production rate $p$ (number of RNA copies/ml/h), the conversion factor between virus produced and virus observed by titration $\rho$ (no of PFU/RNA copies), the rate at which infectious virus lose infectivity $c$ (virus decay rate, 1/h), and a rate of virus particle loss $c_\mathrm{RNA}$. Here $n_E$ and $n_I$ are the numbers of eclipse and respectively infectious compartments and the virus particle production rate per cell, $p_\mathrm{RNA}$ is defined as from \cite{holder}
\begin{equation}
p_\mathrm{RNA} = 0.5 \times 10^6 \times p
\end{equation}
To determine the in vitro infectivity of a particular strain, \cite{holder} used the infecting time
\begin{equation}
t_\mathrm{infect} = \sqrt{\frac{2}{\rho p \beta}}
\label{tinfect}
\end{equation}
which is the amount of time required for a single infectious cell to cause the latent infection of one more, within a completely susceptible population. In the same framework we define the reproductive number $R_0$, the number of cases one case generates on average over the course of its infectious period, in an otherwise uninfected environment:
\begin{equation}
R_0 = \frac{\rho p \beta}{c+c_\mathrm{RNA}}
\label{rzero}
\end{equation}
The model in equation (\ref{re0}) (composed of the time-differential equations $\dot X_k (\theta_i,t)= 0$ with $k$ state variables $X_k$ and constant model parameters $\theta_i$) describes the virus; we would like to test if this model describes a certain data set of discrete time points $x_j(t)$. In doing so we solve the differential equations and fit the solutions to the data; this process makes use of minimizing the error function:

\begin{eqnarray}
\mathrm{SSR} &=& \sum_j\left( X_{0k}f(\theta_i, t_j) - x_j \right)^2 
\label{sfunction}
\end{eqnarray}
or the sum of the squared residuals (SSR) $r_j = X_{0k}f(\theta_i, t_j) - x_j$ with respect to the parameter set $\theta_i$, where $X_k(t_j) = X_{0k}f(\theta_i, t_j) \equiv q_j$ is a solution of the $\dot X_k (\theta_i,t)= 0$ differential system in the vicinity of the $j$th data point and $X_{0k}$ is the state variable $X_k$'s initial value. We see immediately that both the nature of any given minimum of the SSR for a given parameter set and that very same recovered parameter set $\theta_i$ will depend on the initial conditions $X_{0k}$
\begin{equation}
\left (\frac{\partial \mathrm{SSR}}{\partial \theta}\right)_i = 0 \rightarrow \left (\frac{\partial f(\theta, X_{0k})}{\partial \theta}\right)_i = 0
\label{partial}
\end{equation}
Lest to say, initial values represent themselves a field of its own and how they are treated is very important to the final analysis outcome -- for brevity, we will state that $all$ our initial parameter values are to be chosen within physically and biologically motivated intervals; we will update the reader as to our choice of initial values for each of the experiments described here.

Our data $x_j$ is composed of four data sets for which the two measured (explicit) state variables $X_k(t_j)$ are the infectious viral load $V_\mathrm{PFU}=V_\mathrm{PFU}(t_j)$ and the total amount of virus $V_\mathrm{RNA}=V_\mathrm{RNA}(t_j)$. All other state variables (number of cells $T(t)$, number of cells in eclipse phase $E(t)$ etc.) are derived from the system (\ref{re0}) with appropriate physical initial conditions e.g. $T_0=10^6$, $E_0=I_0=D_0=0$. The system's fixed parameters $\theta_i$ are to be estimated using a fitting-to-data algorithm that will sample values for $\theta_i$ from a parameter space with the aim of minimizing the error function (\ref{sfunction}) by means of (\ref{partial}); specifically, our set of parameters 
\begin{equation}
\theta_i \subset [\tau_E, \tau_I, n_E, n_I, \beta, c, c_\mathrm{RNA}, \rho, p]
\label{paramsis}
\end{equation}
The sampling process and minimization of the SSR can be efficiently done by using a Monte Carlo method. The principle behind any given Monte Carlo analysis \cite{sivia,newman} is the Markov Chain (MC) -- a memoryless chain that can be easily represented by a particle jumping through a set of consecutive states towards the state of lowest potential energy; in this process, the particle will retain only the information describing the current and immediately previous states in order to decide if the step is viable or not, all other information is forgotten. The particle will decide to jump states only based on these two pieces of information: if the energy of the next state is lower than the previous one, the particle will jump states. More generally, we want to generate random draws from a target parametric distribution $\Gamma(\theta_i)$. We then identify a way to construct a Markov Chain such that its equilibrium probability distribution is the target distribution $\Gamma$. If we can construct such a chain then we arbitrarily start from some point in the parameter space and iterate the MC many times. Eventually, the draws we generate would appear as if they are coming from our target distribution. We then approximate the quantities of interest (e.g. mean) by taking the sample average of the draws after discarding a few initial draws (``burn-in'' draws) which is the Monte Carlo component. There are several ways to construct Markov Chains (e.g., Gibbs sampler, Metropolis-Hastings algorithm \cite{newman}).

In the absence of noise, given a set of MCMC ``walkers'' (an elementary particle undergoing a ``random walk'', a physical approximation of the exploration of the Markov state space) exploring the parameter space $\theta_i$, the MCMC engine will minimize the residuals $r_j$ $i.e.$ solve the ODE system $\dot X_k (\theta_i,t)= 0$ given by (\ref{re0}), obtain values for the error function SSR from equation (\ref{sfunction}) for $\theta_i$ and the walkers will converge towards a parametric position where the SSR has a minimum (whether it be local or global); in terms of distributions, assuming no major physical noise source affecting the measurements in a systematic manner  and an almost perfect mathematical model across all data sets, residuals $r_j$ are normally distributed and the SSR has a $\chi^2$ distribution with $p$ degrees of freedom. Using the probability distribution function (PDF) of a $\chi^2$-distributed variable and the empirical condition that the lower the SSR the better the model fit to the data, the likelihood in a maximum likelihood estimation case may be given by equation (\ref{likelihood})
\begin{equation}
\ell (\mathrm{SSR}) = C\mathrm{e}^{-\mathrm{SSR}/2}
\label{likelihood}
\end{equation}
where $C$ is a constant for a given $p$ degrees of freedom. By definition, equation (\ref{likelihood}) gives the likelihood of a walker transitioning from a given state, characterised by a certain parameter set, to a current state characterised by a parameter set $\theta_i$ and a fitting error SSR($\theta_i$).

From the Bayesian perspective, there are known and unknown quantities: the known quantity is the data, represented by the number of discrete time points $x_j(t_j)$ (evidence); the unknown quantity is the probability $P(\theta_i|x_j, z)$ -- the posterior distribution of parameters $\theta_i$ given evidence $x_j$ and a model $z$ (e.g. system (\ref{re0})). Using Bayes’ rule, this probability function can be written:

\begin{eqnarray}		
P(\theta_i|x_j,z) &=& \frac{P(\theta_i) P(x_j,z|\theta_i)}{P(x_j,z)} \nonumber \\ 
           &=& \frac{P(x_j,z|\theta_i)P(\theta_i)}{\int_0^1 P(x_j,z|\theta_i)P(\theta_i) \mathrm{d}\theta_i} \nonumber \\
           &=& \frac{\ell (\mathrm{SSR})P(\theta_i)}{\int_0^1 \ell (\mathrm{SSR(\theta_i)}) P(\theta_i) \mathrm{d}\theta_i}
\label{bayesarray}
\end{eqnarray}
Assuming a flat prior $i.e.$ $P(\theta_i)=1$, and a finite constant value for the integral $h_0 = \int_0^1 \ell (\mathrm{SSR(\theta_i)}) P(\theta_i) \mathrm{d}\theta_i$, it follows from equation (\ref{bayesarray})
\begin{eqnarray}
P(\theta_i|x_j,z) &\rightarrow& \nonumber \\
          &=& \frac{\ell (\mathrm{SSR})}{h_0} = A\mathrm{e}^{-\mathrm{SSR}/2}
\end{eqnarray}
or
\begin{equation}
P(\theta_i|x_j,z) \propto \ell (\mathrm{SSR})
\label{posteriors}
\end{equation}
It is easy to understand $P(\theta_i|x_j,z)$ now, as the posterior distribution of parameters $\theta_i$, the very same above mentioned distribution $\Gamma(\theta_i) \equiv P(\theta_i|x_j,z)$ in which the MCMC walkers will $walk$ through $i.e.$ which the MCMC analysis will repeatedly sample to obtain a set of values for the SSR. The walkers will aim towards regions of low SSR (valleys) and there they will explore the parameter space in a more thorough manner (higher sampling rate) until they will reach an SSR minimum, where the sampling will cease and the chain will end. There is one danger to this process: certain walkers may have the tendency, highly dependent on the initial parameter values, to converge to a local minimum that is sometimes rather far from the global minimum. The global minimum needs to be reached since a local minimum can often present us with relatively small fitting errors (SSRs) but a completely different set of parameter values from the true values given by reaching the global minimum. This is why it is necessary to ``aid'' the MCMC walkers to step over these local valleys, by setting a correct analysis statistic, following in the next section.

\section{Weighted analysis statistic}
\label{modelandmethod}
\vspace{0.3cm}

\begin{figure}[ht!]
\includegraphics[scale = 0.3]{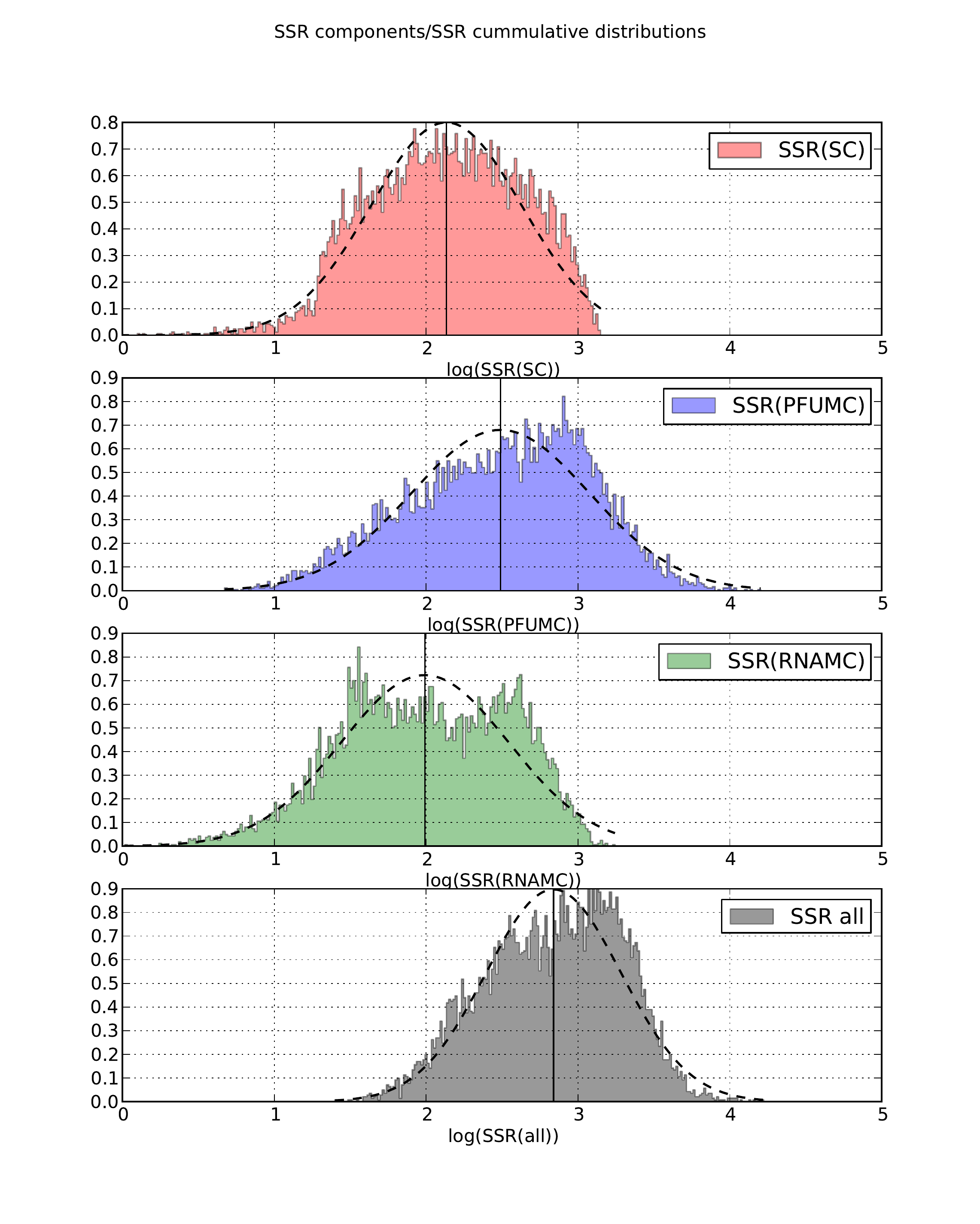}
\caption{SSR components distributions; last panel -- additive overall SSR (the sum of components, equation (\ref{ssrSumski})) distribution; normal distributions are fitted (dashed lines) and their means and variances are shown in Table \ref{ssrTab}.}
\label{ssrdistr}
\end{figure}

Let's introduce now the four independent experiments that we have data for (the data for this work was taken from \cite{holder}, publicly available in the form of viral load time evolution curves); all four data sets are modelled by the same ODE system (\ref{re0}) that uses the same set of parameters $\theta_i$ as presented in \cite{holder} and in equation (\ref{paramsis}). These four experiments are called single-cycle viral load measurement (SC), multiple-cycle viral load measurement (PFUMC), multiple-cycle RNA load measurement (RNAMC) and mock yield viral load measurement (MY); for an in-depth description of each of these experiments and how they were carried, consult \cite{holder}, it is not the purpose of this work to describe the experiments from a methodological point of view since the author has not conducted any of the experiments, nor he had he any involvement in the data collection procedure. It is, however, important to mention that, from a statistical point of view, some parameters are better recovered by fitting the model to only a certain subset of the four data sets than the entire set. In \cite{holder} the same ODE model is applied to $all$ four data sets, and it is assumed that the SSR values have similar distributions across the data sets ($\chi^2$ with equal degrees of freedom $p$ or log-normal distributions with equal means and variances), therefore the $sum$ of the SSR values of the individual SC, PFUMC, RNAMC and MY experiment components was used as overall analysis statistic 
\begin{eqnarray}
\mathrm{SSR} &=& \mathrm{SSR}_\mathrm{SC}+\mathrm{SSR}_\mathrm{PFUMC}+ \nonumber \\
    &+& \mathrm{SSR}_\mathrm{RNAMC}+\mathrm{SSR}_\mathrm{MY}
\label{ssrSumski}
\end{eqnarray}
In reality, due to high levels of experimental noise (that we do not account for here and \cite{holder} does not analyze either) $and$ due to the mathematical model performing in an uneven manner in describing the data sets, these SSR components $do$ $not$ have log-normal distributions but rather are $\chi^2$ distributed with different degrees of freedom $p$. For simplicity, we $will$ approximate these distributions as log-normal but we will apply weights to each SSR component in the following manner: from performing 10,000 fitting trials with parameters randomly sampled from a 10-dimensional parametric ``box'' (see Table \ref{TheBox}), we obtain the SSR component distributions shown in Figure \ref{ssrdistr}. Fitting normal distributions in log-normal space, we obtain a set of reference numbers in Table \ref{ssrTab}.
\begin{table}[ht!]
\begin{tabular}{|l|l|l|}
\hline
\hline
SSR component & Mean & Std. Dev. \\
\hline
SC & 136.0 & 3.2 \\
PFUMC & 308.0 & 4.0 \\
RNAMC & 98.0 & 3.6 \\
MY & 12.7 & 2.9 \\
\hline
Combined & 689.0 & 3.0 \\
\hline
\hline
\end{tabular}
\newline
\caption{Measures of SSR component means and variances, when assuming log-normal distributions, from running 10,000 fitting trials with parameters randomly sampled from the parameter ``box'' in Table \ref{TheBox}.}
\label{ssrTab}
\end{table}
We must stress upon the fact that the log-normal approximation does not play any other role in the likelihood formulation apart from assigning weights to the different SSR components. From Table \ref{ssrTab} we see an approximately twofold contribution from the PFUMC component to the overall SSR sum. This is an $average$ effect, for a given uniformly sampled parameter box with large sampling boundaries, and does not reflect any local behaviour of the SSR components w.r.t. the additive overall SSR (\ref{ssrSumski}).  

%
%

This effect has implications when performing a parameter estimation analysis: if there are no weights applied on the SSR components, on average, the parameter subset that optimally describes the PFUMC data set will always be dominant. 
Poorly constrained parameters, in the case of using an additive unweighed likelihood from four data sets, have a very low recoverability for certain regions of the parameter space where one or more components of the likelihood can not constrain effectively the others. In terms of error analysis, in these regions, the Gaussian-stationary distribution of the residuals (mean zero and variances equal for all four data sets) is not effective any more. 

\begin{figure}[ht]
\includegraphics[scale = 0.35]{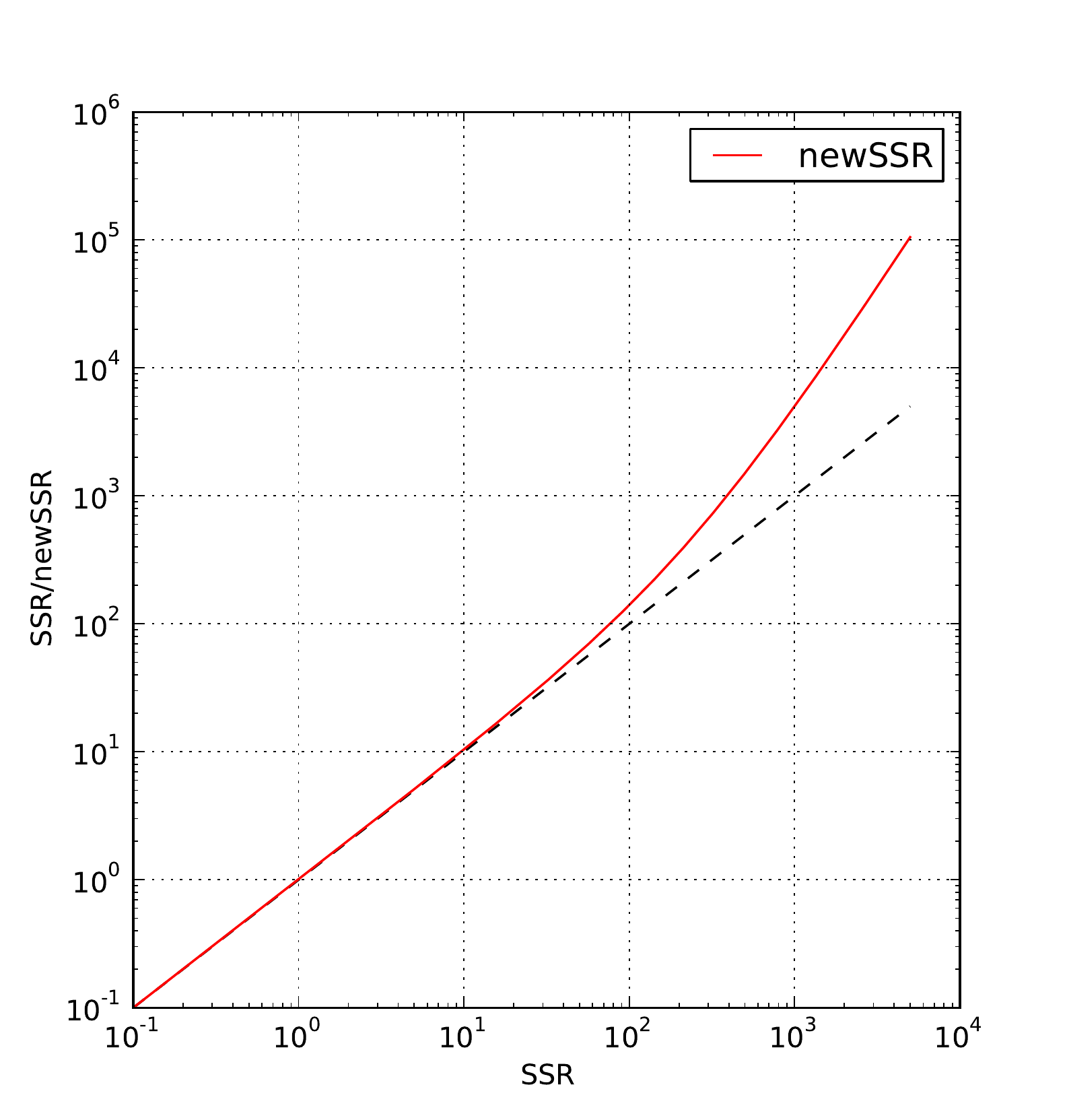}
\caption{New weighted SSR (from equation (\ref{newSSR}), continuous line) compared to standard unweighed additive SSR (from equation (\ref{ssrSumski}), dashed line), as used in previous works as analysis statistic.}
\label{newSSRfig}
\end{figure}

Weighting the SSR components has to be done based on each component's $\chi^2$ distribution parameters (number of effective degrees of freedom), but because it is quite difficult to account for effective degrees of freedom in order to weigh the likelihood components (defined as number of data points minus number of effective model parameters), we could devise a simple method to weigh by component by re-writing the total SSR:
\begin{equation}
\mathrm{SSR}_{\mathrm{new}} = \sum_{a=1}^{a=4} \frac{\mathrm{SSR}_a}{n_a}\left(1+\frac{\mathrm{SSR}_a}{\mathrm{SSR}_{0,a}}\right)
\label{newSSR}
\end{equation}
where $a=4$ is the number of components -- PFUSC, PFUMC, RNAMC and MY, $n_a$ is the number of points per data set $a$, and $\mathrm{SSR}_{0,a}$ is a fixed threshold value for each of the data sets, that aids the MCMC walkers to move rapidly away from a parameter region that yields relatively low SSR values but poor parameter values (a ``steep valley'' of a local minimum). For a fixed $SSR_{0,a}=250$ the new $SSR$ compared to the simple additive SSR, will have the same profile for low values and a much stronger profile for high values, see Figure \ref{newSSRfig}.
Thus, from equation (\ref{newSSR}) the likelihood (\ref{likelihood}) will be rewritten to
\begin{equation}
\ell = C\mathrm{exp}\left(-\sum_{a=1}^{a=4} \frac{\mathrm{SSR}_a}{2n_a}\left(1+\frac{\mathrm{SSR}_a}{\mathrm{SSR}_{0,a}}\right)\right)
\label{newlikelihood}
\end{equation}
By definition, equation (\ref{newlikelihood}) gives the likelihood of a walker transitioning from a given state, characterised by a certain parameter set, to a current state characterised by a parameter set $\theta_i$ and a $weighted$ and $thresholded$ fitting error SSR($\theta_i$). By weighting and thresholding the standard additive SSR and constructing likelihood (\ref{newlikelihood}), we will aid the MCMC walkers to converge faster to the low SSR region and find the global minimum corresponding to min(SSR).

\section{Computational aspects: optimal integrator settings and functional expression of the model}
\label{computation}
\vspace{0.3cm}


Now that we have the four data sets, a model (\ref{re0}) with parameters (\ref{paramsis}) to be estimated, and a likelihood (\ref{newlikelihood}) function, we are ready to use an MCMC engine to obtain posterior distributions (\ref{posteriors}) of our parameters. For an MCMC engine we used $emcee$, presented in \cite{emcee}. $emcee$ is a Python-based, fully-parallelizable, MCMC engine that takes a likelihood expression ((\ref{newlikelihood}) in our case) and a set of initial values to sample the parameter posteriors. $emcee$ is becoming a preferred tool to data analysis from different fields, see \cite{emceeex1,emceeex2} for examples of its practical use.

In order to simulate the full set of MCMC production runs with $emcee$, and since at each iteration, to be able to compute the likelihood, the SSR function (\ref{newSSR}) needs to be computed, we ran a set of 20,000 fitting trials with parameters randomly sampled from a 10-dimensional ``box'' with dimensions listed in Table \ref{TheBox}. The dimensions for this ``box'' were chosen based on phenomenology of the viral data -- the ``box'' is large enough to comprise most possible parameter values hit by the MCMC walkers
\begin{table*}[ht!]
\centering
\begin{tabular}{|a|b|a|}
\hline
\hline
Parameter & Min. value & Max. value \\
\hline
\hline
$p$(1/h) & $\frac{\mathrm{max}(V_\mathrm{PFUMC})^*}{10^4}$ & $10^2 \times$ max($V_\mathrm{PFUMC}$) \\
$\beta$(1/h) & $10^{-3} \times \frac{\mathrm{max}(V_\mathrm{RNAMC})}{\left(\mathrm{max}(V_\mathrm{PFUMC})\right)^2}$ & $10^{3} \times \frac{\mathrm{max}(V_\mathrm{RNAMC})}{\left(\mathrm{max}(V_\mathrm{PFUMC})\right)^2}$ \\
$c$(1/h) & 0.01 & 1.0 \\
$c_{RNA}$(1/h) & 0.001 & 0.1 \\
$\rho$ & $10^{-5}$ & 1.0  \\
$\tau_E$(h) & 3.0 & 30.0 \\
$\tau_I$(h) & 0.1 & 60.0 \\
$V_\mathrm{0MC}^{**}$(PFU/ml) & $\frac{V_\mathrm{0PFUMC}}{10^5}$ & $10^2 \times V_\mathrm{0PFUMC}$ \\
\hline
\hline
\end{tabular}
\newline
* $\mathrm{max}(V_\mathrm{PFUMC})$ represents the maximum viral load from the PFUMC data set;
\newline
** $V_\mathrm{0MC}$ represents the initial value for the PFUMC viral load, parameter to be estimated as well.
\caption{The parametric 10-dimensional ``box'' from wich we randomly sampled parameters for the 20k ODE solving trials.}
\label{TheBox}
\end{table*}

\begin{figure}[ht!]
\includegraphics[scale = 0.35]{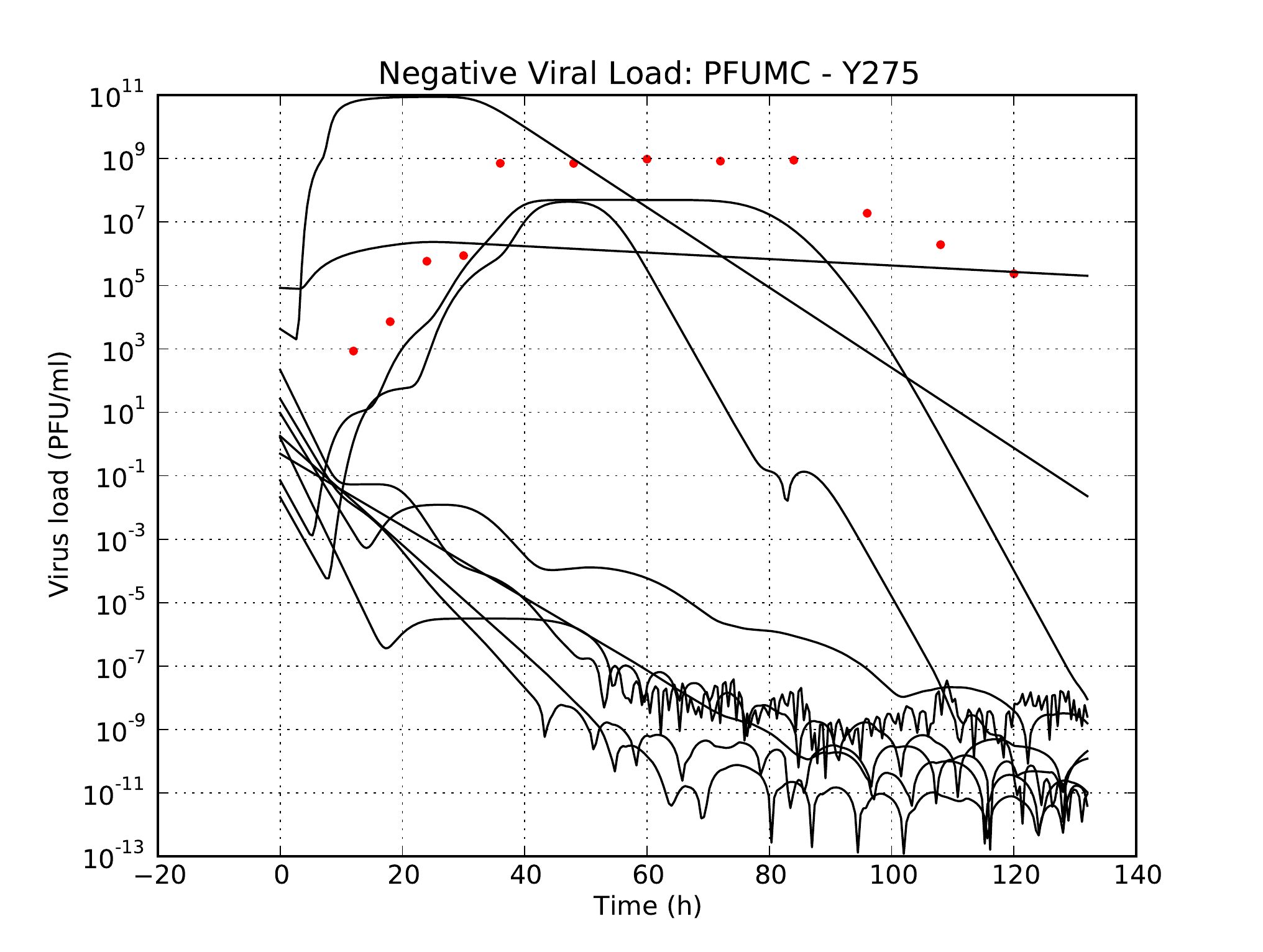}
\newline
\caption{Failed integrations with $scipy.integrate.odeint()$ exiting with Python's ``math domain error'' (solutions obtained using the parameter sets drawn from the ``box'' given in Table \ref{TheBox}; we notice a series of problematic behaviour types of the solutions -- oscillations (with different amplitudes and frequencies) at artificially very low values of viral load $V_\mathrm{PFUMC}(t)$, typical cases of solver's incapacity to reach a stable solution.}
\label{zooCurves}
\end{figure}

As a result of these test runs, we have noticed failure rates of 16-18\% of the ODE solver, a rather significant fraction when performing a full production MCMC analysis. A few of the failed cases are shown in Figure \ref{zooCurves}, where the dots represent the actual data points and the lines represent the solutions offered by the solver. These are consequences of system (\ref{re0}) being a $stiff$ ODE system. In mathematics, a stiff differential equation is an equation for which certain numerical methods for solving it are unstable, unless the step size is taken to be extremely small \cite{odes}. It has been proven difficult to formulate a precise definition of stiffness, but the principal idea is that the equation includes some terms that can lead to rapid variations in the solution. While simulating the MCMC runs, we have noticed the failure of the employed numerical integrator -- Scientific Python's $scipy.integrate.odeint()$ -- to give exact solutions to the ODE system and initially there was no simple rule or pattern to where the integrator failed. These failures are grouped as follows, depending on the type of exit output:
\vspace{0.15cm}
\begin{itemize}
\item
The most frequent exit case of the integration process was associated with Python's ``math domain error'', a typical computing error that occurs when (in our case) the routine tries computing a logarithm of a negative real number -- in this case, the computation of the SSR value was not completed due to negative values of the PFUMC viral load; this error would be recorded for another case as well: the integrator would fail due to the need for a larger number of integration steps (larger than the default maximum of 500) -- in this case the integration would be done up to a certain point where a steep change in the $V(t)$ curve, the integrator would exit and the rest of the viral load values would be of order $10^{-300}$, below the machine precision of $numpy.log10$;  
\item
On occasions, we noticed 'NaN' values of the SSR, produced by positive but infinitesimally small values of viral output, referring us to the previous case;
\end{itemize}
\vspace{0.15cm}

The $scipy.integrate.odeint(f(t),t,y0,*args)$ ODE integrator (see documentation available here \url{http://docs.scipy.org/doc/scipy/reference/generated/scipy.integrate.odeint.html}) uses an adaptive timestep Runge-Kutta 4-5 method written inside its Fortran solver library (see \url{http://docs.scipy.org/doc/numpy/user/install.html} for included compiler instructions) to integrate systems of ordinary differential equations. The number of time steps taken by the integrator varies but is capped by default at $mxstep$=500. Also, the $odeint$ function takes a list of output timesteps, but the Fortran routine is called once for each desired output and uses the previous call as the initial conditions for the next one. This can prove detrimental in cases where the specified time points where the integral should be evaluated is small (order 12-16, as in our data sets). The problem arises from the fact that, unlike more stable but slower MATLAB integrators, $odeint$ does not use $only$ the initial and final time points to generate a smooth, continuous integral and further on use interpolation to get the function values at each time point; it rather integrates piecewise for each segment determined by the specified time points series, process that is prone to breaking if there are too few time points to allow for continuity. One solution to this problem is to increase the parameter $mxstep$ - the number of integration steps -- that is capped by default at 500. Increasing $mxstep$ to 5000 or 10000 would insure that even for order 12-16 data points the integrator will solve the system for all the specified time points, and will not exit without reaching a stable solution.


Unfortunately, increasing the maximum number of integration steps $mxstep$ will affect the time it takes $scipy.odeint$ to solve a system: if the system is solvable with the default maximum $mxstep \leq 500$, by increasing $mxstep$ to, say, 10000, will not increase the solving time, as seen from Table \ref{timing}; increasing $mxstep$ will only increase the integration time for systems that $scipy.odeint$ can not solve for a given number of time points and will exit after the first one or two integration points, most of the cases we are faced with. Table \ref{timing} shows the time it took the solver to integrate 100 times the same ODE system, with the same parameters, same initial conditions, same initial and final time, only with different numbers of integration time points and $mxstep$s.
\begin{table}[ht!]
\begin{tabular}{|l|l|}
\hline
\hline
12 time points & Time/100 runs \\
\hline
mxstep=5e2 & dt= 51.8s \\
mxstep=1e3 & dt= 169.8s \\
mxstep=1e4 & dt= 167.9s \\
mxstep=1e6 & dt= 221.9s \\
\hline
\hline
100 time points & Time/100 runs \\
\hline
mxstep=5e2 & dt= 188.6s \\
mxstep=1e3 & dt= 188.9s \\
mxstep=1e4 & dt= 189.9s \\
mxstep=1e6 & dt= 206.4s \\
\hline
\hline
\end{tabular}
\newline
\caption{Time it took $scipy.odeint()$ to integrate 100 times the same ODE system, with the same parameters, initial conditions, initial and final time, only with different numbers of time points and $mxstep$'s. The reason why for 12 time points and $mxstep$=500 the run time is short is that the integrator $exits$ after the second or third time point with no valid solution.}
\label{timing}
\end{table}
Thus, using a larger $mxstep$ than the system default one is a necessary option for an MCMC production run, but it will, in turn, increase the processing time.

One could apply corrections for accuracy by changing the default values for tolerances $rtol$ and $atol$ (internal ODE solver error handling parameters, see documentation at \url{http://docs.scipy.org/doc/scipy/reference/integrate.html}) to orders of $10^{-30}$. The problem with reducing the tolerance to very small values is that the solving for an otherwise exact within the default tolerance solution will take much longer time (of order 15-20 times longer than the solving without adjusting tolerance) -- this, in light of a set of numerous MCMC simulations, could be a very much unwanted behaviour.

There is, however, an alternative solution to the ODE incapacity problem: by effecting a change of state variable, we rewrite the ODE's from system (\ref{re0}) in an exponential manner. Our state variables are $T,E_i,I_j,V_\mathrm{PFU},V_\mathrm{RNA}$ and $D$ with $i,j$ the number of eclipse and infectious compartments respectively. Consider $\Phi$ any of these real and positive defined state variables. We will operate a change of variable that will map $\Phi$ to a different real (both positive and negative defined) variable $\Omega$ with the following functional relation:
\begin{equation}
\Phi \rightarrow \Phi:=\mathrm{e}^{\Omega}
\end{equation}
and this way the differential forms, by applying the chain rule, will be:
\begin{eqnarray}
\frac{\mathrm{d}\Phi}{\mathrm{d}t} &=& \left(\frac{\partial \Phi}{\partial \Omega}\right)_{t}\left(\frac{\partial \Omega}{\partial t}\right)_{\Phi} \nonumber \\
                                   &\equiv& \Phi \frac{\mathrm{d}\Omega}{\mathrm{d}t}
\end{eqnarray}
With this in mind we rewrite the computational form of our ODE system (\ref{re0}) as:
\begin{eqnarray}
\dot V_\mathrm{PFU} &=& \rho p \mathrm{e}^{-V_\mathrm{PFU}}\left(\sum_{j=1}^{j=n_I}\mathrm{e}^{I_j}\right) - (c+c_\mathrm{RNA}) \nonumber \\
\dot V_\mathrm{RNA} &=& p \mathrm{e}^{-V_\mathrm{RNA}}\left(\sum_{j=1}^{j=n_I}\mathrm{e}^{I_j}\right) - c_\mathrm{RNA} \nonumber \\
\dot T &=& -\beta \mathrm{e}^{V_\mathrm{PFU}} \nonumber \\
\dot E_1 &=& \beta \mathrm{e}^{\left(T+V_\mathrm{PFU}-E_0\right)} - k \nonumber \\
\dot E_i &=& k \mathrm{e}^{-\Delta E} - k \nonumber \\
\dot I_1 &=& k \mathrm{e}^{E_{n_E}-I_0} - \delta \nonumber \\
\dot I_j &=& \delta \mathrm{e}^{-\Delta I} - \delta \nonumber \\
\dot D &=& \delta \mathrm{e}^{I_{n_I}} \mathrm{e}^{-D}
\label{re}
\end{eqnarray}
with $E_0$ and $I_0$ the values of $E$ and $I$ in the first eclipse and infectious compartments respectively and $\Delta E=E_{i} - E_{i-1}$ and $\Delta I=I_{j} - I_{j-1}$.


The integration initial conditions are straightforward from (\ref{re0}) and (\ref{re}): natural logarithms of the original, non-exponential ODE system, $i.e.$ $V^{(\ref{re})}_\mathrm{PFU}=\mathrm{ln}(V^{(\ref{re0})}_\mathrm{PFU}), V^{(\ref{re})}_\mathrm{RNA}=\mathrm{ln}(V^{(\ref{re0})}_\mathrm{RNA}),T^{(\ref{re})}=0.0, E^{(\ref{re})}=I^{(\ref{re})}=D^{(\ref{re})}=-n$ where $n$ is a relatively large natural number (30-100). The reason why $E^{(\ref{re})}=I^{(\ref{re})}=D^{(\ref{re})}=-n$ is that we need a good approximation for null initial conditions from $E^{(\ref{re0})}$ and $I^{(\ref{re0})}$ and these do not introduce errors in the solver due to too small a number.

\begin{figure}[ht!]
\begin{minipage}[b]{0.3\linewidth}
\centering
\includegraphics[scale = 0.19]{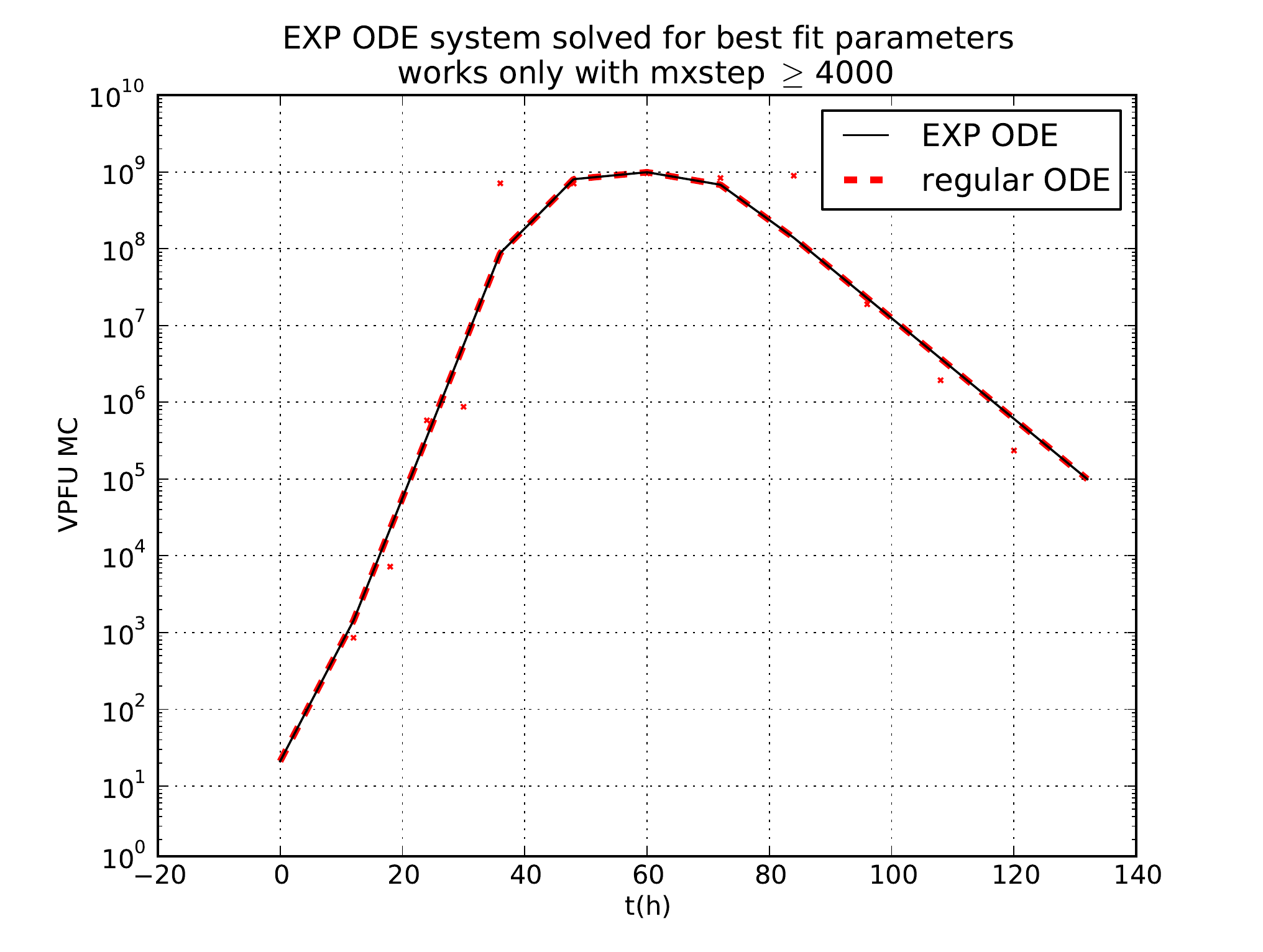}
\end{minipage}
\hspace{1.0cm}
\begin{minipage}[b]{0.3\linewidth}
\centering
\includegraphics[scale = 0.19]{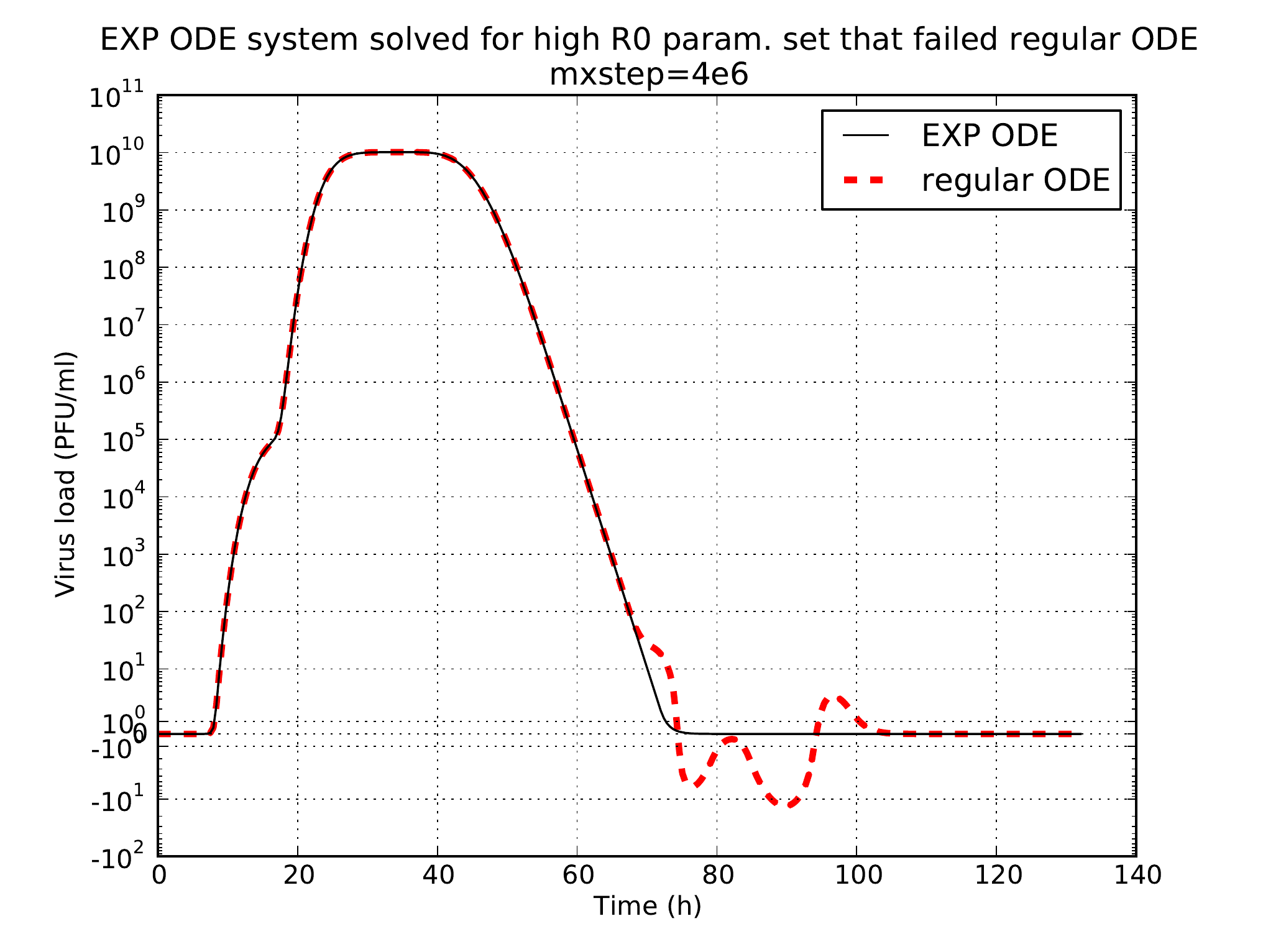}
\end{minipage}
\newline
\caption{(left) shows system \ref{re} finds the exact same solution for a favourable best fit parameter set (the SSR was identical between the two systems (\ref{re0}) and (\ref{re}) to precision order $10^{-5}$); the system (\ref{re}) is more computationally intensive and needs longer time to solve than the non-exponentiated system (\ref{re0}); it also needs a larger number of the maximum step value (we have obtained the best fit curve with $mxstep \geq 4000$). The time it took system \ref{re} to integrate with best fit parameters was 48.9s (for 50 trials) whereas the non-exponentiated system (\ref{re0}) took 29.6s for the same number of trials. (right) shows the same comparison but with a set of unfavourable parameters that would induce the non-exponentiated system (\ref{re0}) to produce numerically unstable solutions (hence damped oscillatory behaviour); system (\ref{re}) produces the correct solution and in a shorter time - 55.3s (for 50 trials) whereas the non-exponentiated system took 159.2s for the same number of trials, with a high percentage of unstable solutions}
\label{bf}
\end{figure}

We have tested the re-written ODE system (\ref{re}) against the one already in use (\ref{re0}) in \cite{holder}. Figure \ref{bf} (left) shows system (\ref{re}) finds the exact same solution for a favourable best fit parameter set (the SSR was identical between the two systems (\ref{re0}) and (\ref{re}) to precision order $10^{-5}$); the system (\ref{re}) is more computationally intensive and needs longer time to solve than the non-exponentiated system (\ref{re0}); it also needs a larger number of the maximum step value (we have obtained the best fit curve with $mxstep \geq 4000$). The time it took system (\ref{re}) to integrate with best fit parameters was 48.9s (for 50 trials) whereas the non-exponentiated system (\ref{re0}) took 29.6s for the same number of trials. Figure \ref{bf} (right) shows the same comparison but with a set of unfavourable parameters that would induce the non-exponentiated system (\ref{re0}) to produce numerically unstable solutions (hence damped oscillatory behaviour); system (\ref{re}) produces the correct solution and in a shorter time - 55.3s (for 50 trials) whereas the non-exponentiated system took 159.2s for the same number of trials, with a high percentage of unstable solutions.

We repeated the 20,000 trials with parameters randomly sampled from the ``box'' of Table \ref{TheBox} above but using the system given by equation (\ref{re}) this time. $All$ solutions were stable, so failure rate in computing a numerical value for the SSR due to non-physical negative $V(t)$ values is 0\%.

\section{Analysis and results}
\label{results}
\vspace{0.3cm}

\begin{figure}[ht!]
\begin{minipage}[b]{0.3\linewidth}
\centering
\includegraphics[scale = 0.13]{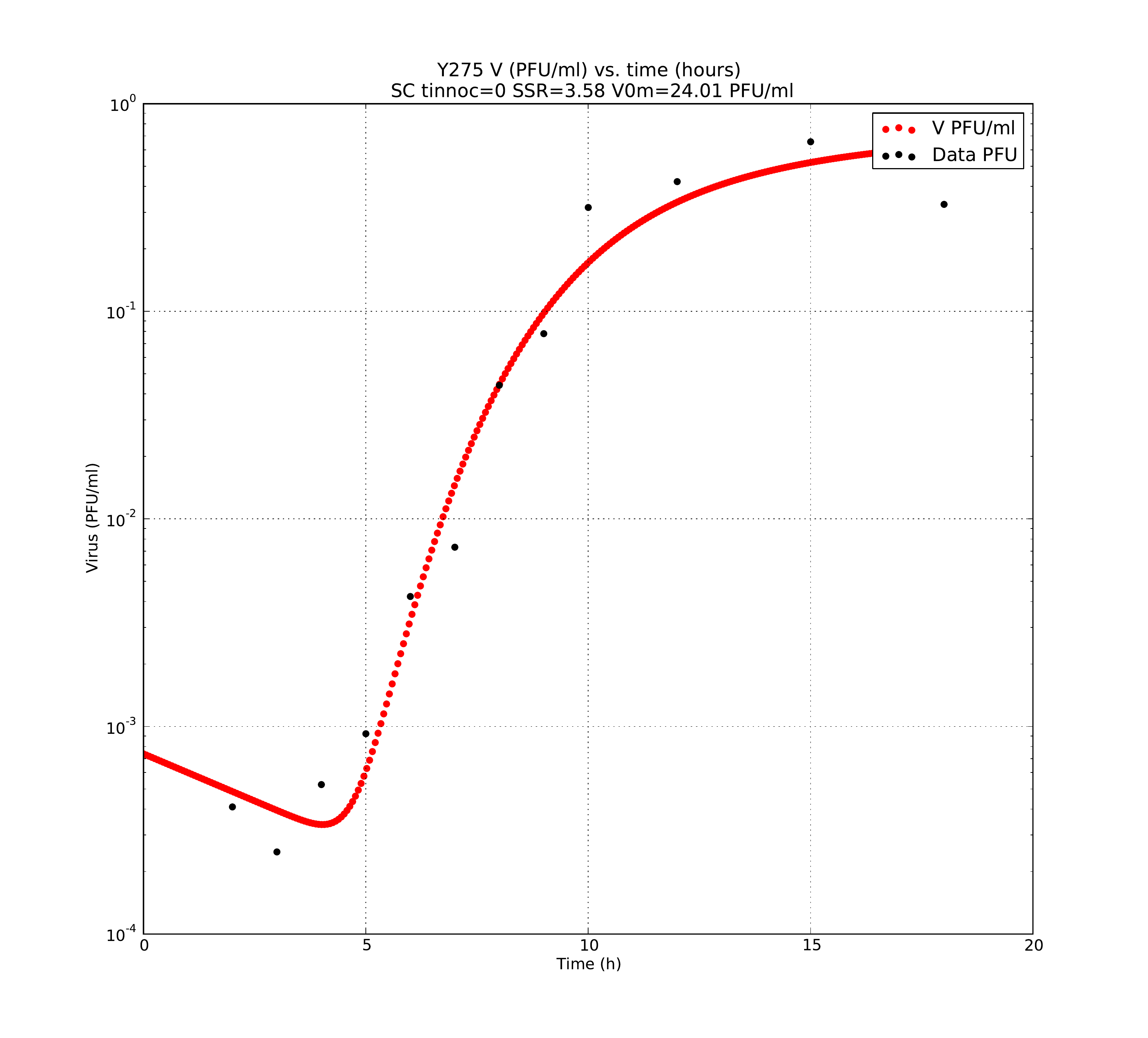}
\end{minipage}
\hspace{1.0cm}
\begin{minipage}[b]{0.3\linewidth}
\centering
\includegraphics[scale = 0.13]{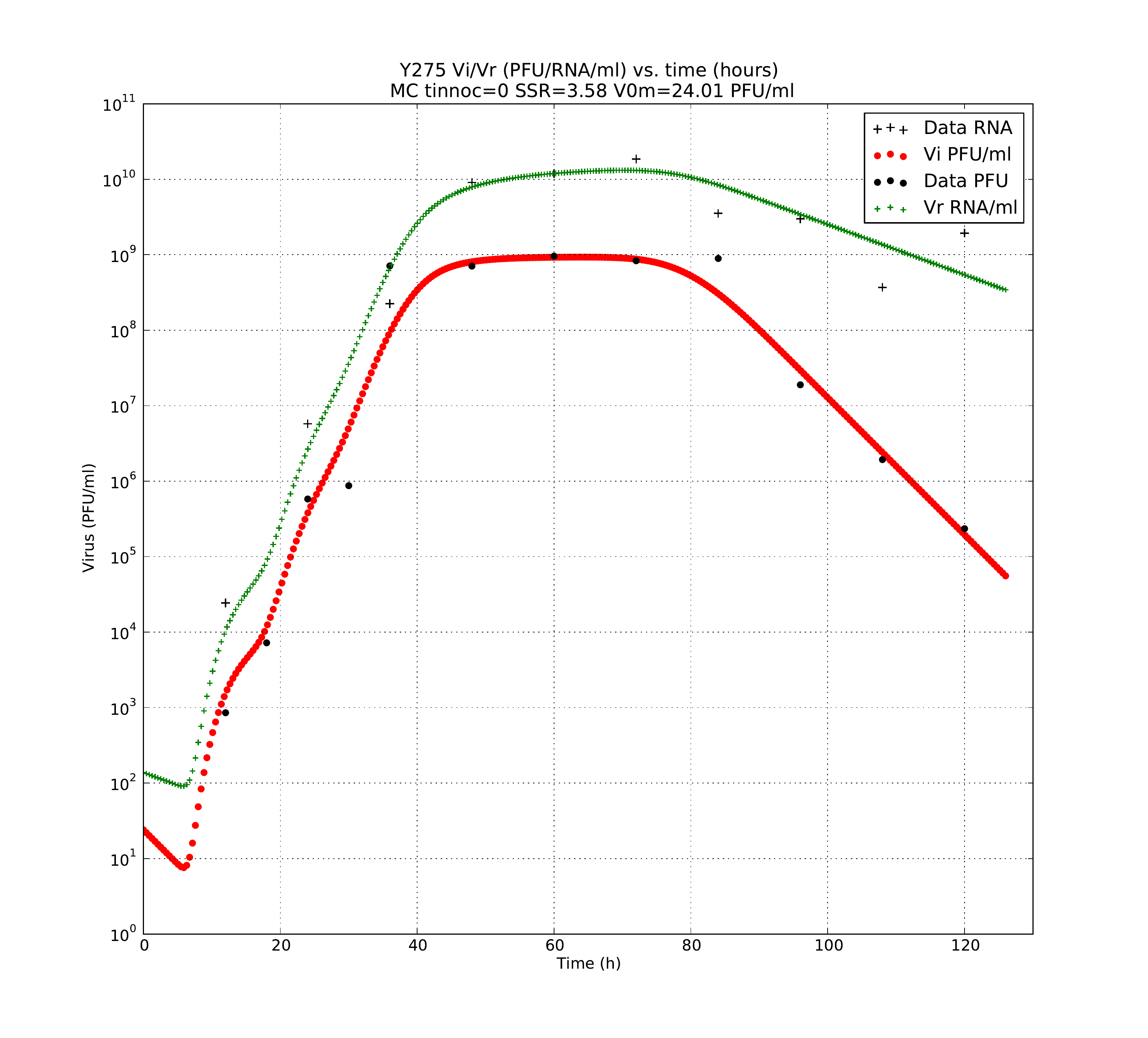}
\end{minipage}
\newline
\caption{``Best fit'' curves of model (\ref{re}) for the SC, PFUMC and RNAMC data sets; parameters for these curves have been used as centers for the intervals from which the MCMC walkers have been started from, as their initial conditions.}
\label{BestFits}
\end{figure}

In order to obtain a set of ``best fit'' parameters $i.e.$ a set of parameters for which model (\ref{re}) best describes the data, we performed a number of $\sim$1500 bootstrap runs. Any given bootstrap run is characterised by a set of initial conditions (initial parameter values, randomly chosen from the parametric ``box'' (Table \ref{TheBox}), an ODE system (\ref{re}) and an error function that quantifies how well the solution fits the data, given in equation (\ref{newSSR}); for the thresholds we chose values of order 3 times larger than the ``best fit'' values $i.e.$ $\mathrm{SSR}_a$ in equation (\ref{newSSR}) we chose $\mathrm{SSR}_\mathrm{SC}=10$, $\mathrm{SSR}_\mathrm{PFUMC}=8$, $\mathrm{SSR}_\mathrm{RNAMC}=6$ and $\mathrm{SSR}_\mathrm{MY}=3$ ($\ddagger$). This ensemble is passed to a least squares engine that walks the system through a large combination of parameter values, with each iteration coming closer to the global minimum position in parameter space $i.e.$ minimizing the SSR. We then chose the single parametric combination with the lowest SSR from these bootstrap runs. Such a global minimum ``best fit'' to the data is shown in Figure \ref{BestFits} for three of our four data sets (SC, PFUMC and RNAMC); the parameters for this case are used to construct the intervals to start the MCMC walkers from: the ``best fit'' parameter set is used as the center of the interval with boundaries given by the mean of all the bootstrap parameter values.

The bootstrapping process randomly samples the fitting errors with replacement; this set is usually assumed to be from an independent and identically distributed population, and it is implemented by constructing a number of resamples with replacement of the same size as the initial set. The main problem of using bootstrapping to estimate biological parameters is that the mathematical model errors are not sampled from the same distribution - the model is applied to different data sets that it describes differently; there is also the experimental noise that introduces fluctuations in the fitting errors, but this we can not model in this work. In our situation, using bootstrapping to estimate parametric confidence intervals will introduce a bias towards the data set(s) that are worse described by the model and the parametres will have broader distributions, usually with heavy tails in the regions where the model constrains poorly those data sets. On the other hand, the MCMC analysis will not assume $all$ error samples are drawn from the same distribution and will most often reject those parameter regions producing the bootstrap distribution tails. This bing said, using bootstrapping is a good way to obtain ``best fit'' parameter intervals that can be further used as initial values for the MCMC runs.

The MCMC simulations have been run on multiple CPU's using the OpenMPI (Open Message Passing Interface \url{http://www.open-mpi.org/}) implementation in $emcee$ package (see \cite{emcee} and the guide at \cite{emceeguide} for instructions on how this mode is configured, specifically MP4Py); the computational load has been divided and managed on Compute Canada's SHARCNET academic and scientific-usage computer clusters (\url{https://www.sharcnet.ca/}). The analysis procedure comprised the following specifications:
\vspace{0.15cm}
\begin{itemize}
\item
According to $emcee$'s user manual \cite{emceeguide} it is desired to start each of the walkers from a favourable position in the parameter space -- as such, we started all the walkers on a tight ball centered at the ``best fit'' parameters of each of the four experiments. The ``best fit'' parameters are seen in Table \ref{resultsTable} together with the walkers' start intervals, labelled as $^{(1)}$ columns;
\item
We ran 2000 iterations for each of the 100 walkers per combined data sets (SC, PFUMC, RNAMC and MY) with 200 iterations for burn-in; sampling was performed in linear parameter space. The $emcee$ package needs just a few inputs to be able to handle the MCMC simulations, the most important input being the an expression for the likelihood of accepting or rejecting the parameter set of any given walker's position in the parameter space. This likelihood is given by equation (\ref{newlikelihood}) and is directly coded in, making use of the SSR thresholds ($\ddagger$);
\item
Chains had average acceptance rates of 30-40\%; convergence and chain mixing was checked using a Gelman-Rubin diagnostics test and chains that did not reach convergence were discarded (the Gelman-Rubin test \cite{gelman} is used to check the convergence of multiple MCMC chains run in parallel; it compares the within-chain variance to the between-chain variance);
\item
Only parameter sets from walkers with SSR<5.0 were kept for final parameter estimation, the rest of the sets being discarded.
\end{itemize}
\vspace{0.15cm}

\begin{table*}[ht!]
\centering
WT-H275 and MUT-H275Y data: $emcee$ MCMC Simulations Results for ODE Model (\ref{re})
\newline
$^{(1)}$ - initial value interval; $^{(2)}$ - obtained median value; $^{(3)}$ - obtained credible interval from posterior distribution
\newline
\begin{tabular}{|b|a|b|b|a|b|b|}


\hline
\hline
Parameter & $^{(1)}$WT-H275 & $^{(2)}$WT-H275 & $^{(3)}$WT-H275 & $^{(1)}$MUT-H275Y & $^{(2)}$MUT-H275Y & $^{(3)}$MUT-H275Y \\
\hline
\hline
Mean eclipse period, $\tau_E$ (h) & 6.6$\pm$3.0 & 6.6 & 6.3--7.1 & 9.1$\pm$3.0 & 9.3 & 9.1--9.6 \\
Eclipse period SD, $\sigma_E$ (h) & 1.2$\pm$2.0 & 1.3 & 1.1--1.5 & 1.6$\pm$2.0 & 1.6 & 1.5--1.8 \\
Infecting time, $t_{\mathrm{infect}}~~^1$ (min) & 31.0$\pm$15.0 & 22.2 & 19.7--23.8 & 22.0$\pm$15.0 & 18.6 & 17.8--19.8 \\
Infectious life span, $\tau_I$ (h) & 49.0$\pm$15.0 & 35.2 & 34.6--36.2 & 41.0$\pm$15.0 & 32.0 & 31.4--32.3 \\
Depletion rate, $c$ (h$^{-1}$) & 0.1$\pm$0.05 & 0.078 & 0.069--0.086 & 0.1$\pm$0.05 & 0.087 & 0.076--0.097 \\
Production rate/cell, $p_{\mathrm{RNA}}$ (RNA copies/h) & 2,200$\pm$1,000 & 3,355 & 3,071--3,700 & 370$\pm$1,000 & 459 & 425--498 \\
Viral burst size, $b~~^2$ (RNA copies) & 110$\pm$50 & 118 & 109--131 & 15$\pm$50 & 15 & 13--16 \\
Reproductive number, $R_0~~^3$ & 2000$\pm$1500 & 3582 & 3101--4085 & 3000$\pm$1500 & 4243 & 3906--4572 \\
\hline
\hline
\end{tabular}
\newline
\newline
$^1~$Computed using equation (\ref{tinfect});
$^2~$Burst size $b = p_{\mathrm{RNA}} \times \tau_I$;
$^3~$Computed using equation (\ref{rzero}).
\newline
\caption{Viral infection parameters obtained from MCMC simulations with the $emcee$ engine: initial values, median values, credible intervals obtained from posterior distributions for both the wild type WT-H275 and the mutant MUT-H275Y H1N1pdm09 strains.}
\label{resultsTable}
\end{table*}

\begin{figure*}[ht!]
\includegraphics[scale = 0.5]{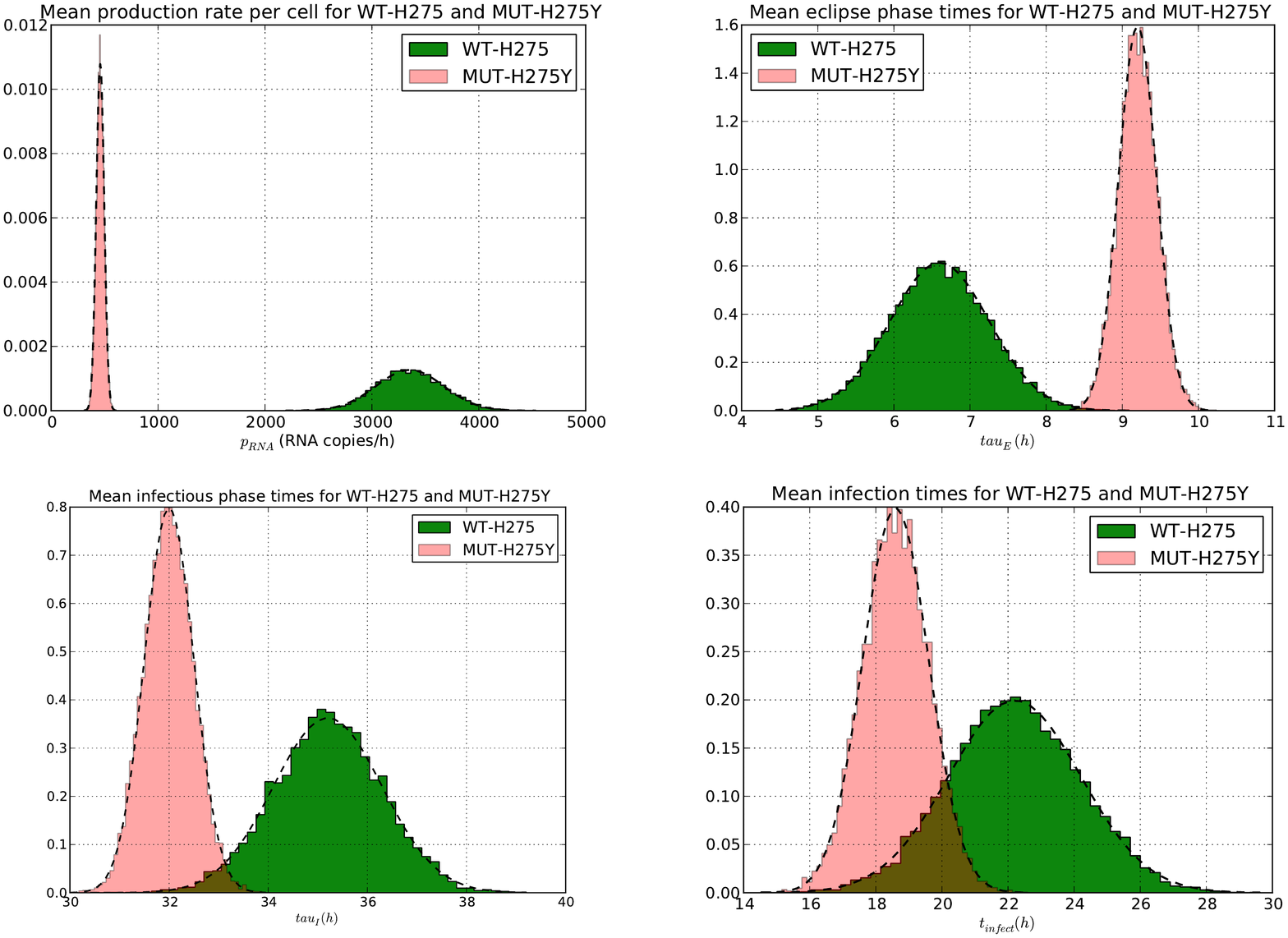}
\newline
\caption{Posterior distributions for four viral parameters: production rate/cell, $p_{\mathrm{RNA}}$ (RNA copies/h, top left corner), mean eclipse period, $\tau_E$ (h, top right corner), infectious life span, $\tau_I$ (h, bottom left corner) and infecting time, $t_{\mathrm{infect}}$ (min, bottom right corner) for both WT-H275 and MUT-H275Y strain data; these distributions have been obtained by running a set of MCMC simulations ($emcee$ engine, 100 walkers, 2000 iterations) and Gaussian distributions have been fitted over (dashed lines) -- the parameters of the fitted distributions are listed in Table \ref{resultsTable}.}
\label{GaussFit}
\end{figure*}

The numerical results are presented in Table \ref{resultsTable}; the listed parameter value $^{(2)}$ is the mean of posterior distributions obtained from the MCMC simulations; the credible interval $^{(3)}$ is obtained from fitting a Gaussian distribution to the posterior distribution, see Figure \ref{GaussFit}. The Gaussian fit agrees very well with the actual posterior distributions obtained from the MCMC simulations.

In order to quantify the difference between the two strains, WT-H275 and MUT-H275Y, from a parametric point of view, we construct the distance in parameter space $\Gamma$
\begin{equation}
\lambda_{\Gamma} = \sqrt{\frac{1}{N}\sum_{i=1}^{i=N} \left(\frac{\mathrm{KS}_i}{w_i}\right)^2}
\label{KSdist}
\end{equation}
where $N$ is the number of estimated parameters, $KS_i$ is the result of a two-sided Kolmogorov-Smirnoff test (see documentation at \url{http://docs.scipy.org/doc/scipy-0.14.0/reference/generated/scipy.stats.ks_2samp.html}) with inputs the two $i$th parameter distributions for WT-H275 and MUT-H275Y (the null distribution of this statistic is calculated under the null hypothesis that the samples are drawn from the same distribution; this test is used to determine how ``similar'' two parametric distributions are, assuming they are normal distributions; the choice of use of a frequentist test over a Byesian one is that there already exists a $scipy$ simple-to-use package that can be called with ease when constructing a parameter-estimation package) and $w_i$ is a set of weights that are chosen function of the recoverability of $i$th parameter (for simplicity we use $w_i=1$ here). We apply a two-sided Kolmogorov-Smirnoff test using the fitted Gaussian distributions. The $KS_i$ test results are listed in Table \ref{ks}. The statistic $\lambda_{\Gamma}$ is a percentage that tells us how different the two compared strains are, in our case, from the MCMC estimated four parameters in Table \ref{ks}, the two strains are 94\% different. Of course, this rationale could be extended for a larger set of parameters, whichever are deemed as significant in the case of comparison, here we are using equation (\ref{KSdist}) with four parameters as an example only.
\begin{table}[ht!]
\centering
WT-H275 and MUT-H275Y data: Kolmogorov-Smirnoff test results
\newline
From equation (\ref{KSdist}), $\lambda_{\Gamma}$ = 0.94\%
\newline
\begin{tabular}{|l|l|}
\hline
\hline
Parameter & $KS_i$ (\%) \\
\hline
\hline
Mean eclipse period, $\tau_E$ & 0.99 \\
Infecting time, $t_{\mathrm{infect}}$ & 0.79 \\
Infectious life span, $\tau_I$ & 0.96 \\
Production rate/cell, $p_{\mathrm{RNA}}$ & 1.00 \\
\hline
\hline
\end{tabular}
\newline
\newline
\caption{Kolmogorov-Smirnoff test results for four (example) parameters.}
\label{ks}
\end{table}

\section{Conclusions}
\label{conclusions}
\vspace{0.3cm}

We used the data from \cite{holder} in a bid to refine the results presented there and formulate new concepts to analytically compare virus strains; we used a different analysis method to estimate the viral dynamics parameters of model (\ref{re0}) -- Markov Chain Monte Carlo simulations. We modified the model to a more computationally-robust model (\ref{re}), formulated a likelihood (\ref{newlikelihood}) and used an existing MCMC computer package ($emcee$ presented in \cite{emcee}) to obtain parameter distributions (posteriors). The MCMC results offer much narrower credible intervals than bootstrapping 95\% confidence intervals (see \cite{holder} for a result using bootstrap replicates); they also offer true parameter probability distribution functions (PDFs) in the form of the posterior distributions: full parameter results can be found in Table \ref{resultsTable} and distributions are shown in Figure \ref{GaussFit}. We used these distributions and fitted Gaussian distributions to extract the credible intervals, the Gaussian fit being rather good. In a novel approach, we used a frequentist two-sided Kolmogorov-Smirnoff test to compare the two viral strains, comparison yielding a 94\% difference from a parametric point of view. By computing the reproductive number $R_0$ we show that the infectivity of the mutant strain is higher than the wid type; this has been shown in \cite{holder} as well but our values for $R_0$ differ and suggest a smaller difference in infectivity between the two strains; this is supported by narrow credible intervals for $R_0$, see \ref{resultsTable}. 

Assuming a flat prior, the MCMC analysis method we present here does not make use of any assumptions with regards to data, and, albeit the data is very small (order 60 data points in total), the results are not only consistent with the ones presented in \cite{holder} but parametric credible intervals are narrower and we could obtain an analytic measure to distinguish the two H1N1pdm09 strains. The use of $emcee$ \cite{emcee} is very easy and, if correctly configured in multi-processor mode, the program runs very fast -- we could obtain the results we present here in a matter of one day. If this analysis method is correctly packaged, one could obtain this type of results in a matter of hours. Using a generic mathematical model and this analysis package could lead to results being obtained very fast, matter that might be very useful in the future in the case of a major pandemic.


\begin{backmatter}



\section*{Acknowledgements}
This work was supported by STFC grant No. ST/L000342/1. The author would also like to thank Drs. Tinevimbo Shiri, Eric Paradis and Eric da Silva (Ryerson University, Toronto, Canada) for the numerous discussions and very useful suggestions they offered him. Many thanks also go to the members of the Gravitational Waves group at Cardiff University, UK, whose member the author is, for support and very useful suggestions. Special thanks go to the author's mom that cooked the best food in the world while writing this article on the occasion of a visit home.

\bibliographystyle{bmc-mathphys} 
\bibliography{bmc_article}      






\end{backmatter}
\end{document}